\documentclass{emulateapj}

\newcommand{\EXO}{\mbox{EXO 0748-676}}

\newcommand{\Msun}{\ensuremath{M_\odot}}

\shorttitle{Neutron Star Line Transport} 
\shortauthors{Chang, Bildsten \& Wasserman}

\begin{document}

\title{Formation of Resonant Atomic Lines during Thermonuclear Flashes
on Neutron Stars} 

\author{Philip Chang} 
\affil{Department of Physics,
Broida Hall, University of California, Santa Barbara, CA 93106;
pchang@physics.ucsb.edu} 
\author{Lars Bildsten} 
\affil{Kavli Institute for Theoretical Physics and Department of
Physics, Kohn Hall, University of California, Santa Barbara, CA
93106, USA; email: bildsten@kitp.ucsb.edu}
\and 
\author{Ira Wasserman}
\affil{Center for Radiophysics and Space Research, Cornell University,
Ithaca, NY 14853; ira@astro.cornell.edu}

\begin{abstract}

Motivated by the measurement of redshifted Fe H$\alpha$ lines during type I
X-ray bursts on EXO~0748-676 (Cottam, Paerels \& Mendez), we study the
formation of atomic Fe lines above the photosphere of a bursting neutron star
($k_BT_{\rm eff} \approx 1-2\,{\rm keV}$).  We discuss the effects of Stark
broadening, resonant scattering and NLTE (level population) on the formation of
hydrogenic Fe H$\alpha$, Ly$\alpha$ and P$\alpha$ lines.  From the observed
equivalent width of the Fe H$\alpha$ line, we find an implied Fe column of $1-3
\times 10^{20}\,{\rm cm}^{-2}$, which is 3-10 times larger than the Fe column
calculated from the accretion/spallation model of Bildsten, Chang \& Paerels.
We also estimate that the implied Fe column is about a factor of 2-3 larger
than a uniform solar metallicity atmosphere.  We discuss the effects of
rotational broadening and find that the rotation rate of \EXO\ must be slow, as
confirmed by the recent measurement of a 45 Hz burst oscillation by Villarreal
\& Strohmayer.  We also show that the Fe Ly$\alpha$ EW $\approx$ 15-20 eV
(redshifted 11-15 eV) and the P$\alpha$ EW $\approx$ 4-7 eV (redshifted 3-5 eV)
when the H$\alpha$ EW is 10 eV (redshifted 8 eV).  The Ly$\alpha$ line is
rotationally broadened to a depth of $\approx 10\%$, making it difficult to
observe with {\it Chandra}. We also show that radiative levitation can likely
support the Fe column needed to explain the line.

\end{abstract}


\section{Introduction}

The recent observation of gravitationally redshifted atomic transition
lines during type I X-ray bursts on the low-mass X-ray binary,
EXO~0748-676 (Cottam, Paerels \& Mendez 2002, hereafter CPM) opens a
new window on neutron star (NS) structure. Using 335 ks of XMM-Newton
observations during its calibration and commissioning phase, CPM
collected 3.2 ks of Type I bursts which was split into a high
temperature part ($k_BT_{\rm BB} \approx 1.8\,{\rm keV}$) and a low
temperature part ($k_BT_{\rm BB} < 1.5\,{\rm keV}$).  In the high
temperature part they identified the H$\alpha$ ($n=2\to 3$) transition
of hydrogenic {\rm }Fe with a redshift of z=0.35.  The same transition
was identified for He-like Fe with the same redshift in the low
temperature part.  Though this is not the first claim of an absorption
feature during type I X-ray bursts, previous Tenma and EXOSAT
observations of a 4.1 keV absorption feature on 2S 1636-536 (Waki et
al. 1984; Turner \& Breedon 1984), EXO 1747-214 (Magnier et al. 1989)
and X1608-52 (Nakamura, Inoue \& Tanaka 1988) were never confirmed
with more sensitive instruments.  CPM's observation is unique in that
two lines in distinct states of the type I burst with the same
redshift were observed, implying that we are seeing the NS surface.

In our previous work (Bildsten, Chang \& Paerels 2003, hereafter BCP),
we explained how Fe can be present above the accreting NS photosphere.
The larger entropy of the photosphere forbids convective mixing from
the deeper burning regions (Joss 1977), making it unlikely that Fe is
dredged up.  In addition, the strong surface gravity rapidly depletes
Fe from the upper atmosphere via sedimentation\footnote{Radiative
levitation is a possible mechanism by which the Fe is suspended (BCP),
an issue we will address in \S~\ref{sec:rad_levitation}.}.  However,
continual accretion during the burst at fairly modest rates ($\dot{M}
> 10^{-12}\,M_{\odot}\,{\rm yr}^{-1}$) can overcome Fe sedimentation.
For \EXO, the persistent flux implies $\dot M> 2\times 10^{-10}
M_\odot {\rm yr^{-1}}$ (Gottwald et al. 1986; CPM) and the
sub-Eddington burst luminosity is unlikely to halt the accretion flow
(see observations of GX 17+2; Kuulkers et al. 2002).

If accretion is the source of the Fe that generates the absorption
line, then accretion must be spread over a large fraction of the star.
Since the NS surface is inside the last stable orbit for the implied
radius of $R=4.4GM/c^2$ (CPM), the protons may reach the NS surface
with a kinetic energy of $\approx 200-300$ MeV/nucleon, and
decelerate via Coulomb scattering (Zel'dovich \& Shakura 1969).  This
filters ions by their charge and mass, stopping incident Fe above the
photosphere at a column of $\approx 1\,{\rm g\,cm}^{-2}$ (Bildsten,
Salpeter \& Wasserman 1992).  The Fe remains there until it is either
destroyed by incident protons (for $\dot{M}>10^{-13}\,{\rm
M}_{\odot}\,{\rm yr}^{-1}$) or sediments out.  The deposition and
nuclear destruction of Fe both depend linearly on $\dot{M}$ and hence
the steady state Fe column is independent of $\dot{M}$ when nuclear
destruction dominates.  For accretion of solar metallicity material,
$N_{\rm Fe} \approx 3.4\times 10^{19} \ {\rm cm^{-2}}$ (BCP).  While
this Fe column is similar to a uniformly abundant solar metallicity
atmosphere ($N_{\rm Fe}\approx 5\times 10^{19}\ {\rm cm^{-2}}$), the
accretion/spallation scenario also predicts a plethora of elements
with $Z<26$ produced by proton spallation of Fe (BCP, Table 1).

In this paper, we discuss the physics of line transfer through a thin
Fe layer above the continuum photosphere. Previous theoretical work on
the formation of spectral features focused on the Fe Ly$\alpha$ line
and edge (Foster et al. 1987; Day, Fabian \& Ross 1992).  However,
these calculations presumed LTE and uniform abundance. In
\S~\ref{sec:model_atmosphere} we show that LTE is a poor assumption as
the radiation field dominates the level population and ionization
balance (London, Taam \& Howard 1986; BCP).  We also discuss the line
broadening physics and show that Stark broadening dominates for Fe
H$\alpha$.  We calculate in \S~\ref{sec:line_formation_calculations}
the Fe column needed to produce the observed H$\alpha$ line and
compare it to the Fe column from accretion/spallation. We also
estimate the equivalent width of the resulting Ly$\alpha$ and
P$\alpha$ transition, and constrain the rotation rate of the NS
from the observed linewidth. We close by estimating the Fe column that
might be sustained by radiative levitation, and summarize our results
in \S~\ref{sec:conclusion}.

\section{Model Atmospheres}\label{sec:model_atmosphere}

\begin{figure}
\plotone{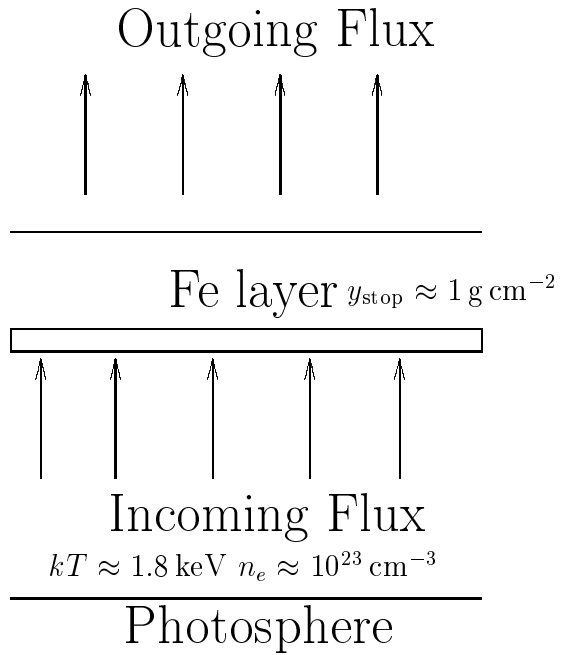}
  \caption{Schematic model of the line forming region in a NS
  atmosphere during a Type I burst.}
  \label{fig:model}
\end{figure}

We model the thin Fe line forming region as a one-zone scattering
layer (Mihalas 1978) at fixed temperature and density that is
illuminated from below by the continuum photosphere (see Figure
\ref{fig:model}).  We consider radiation and gas temperatures of
$k_BT_g \approx k_BT_{\rm eff} \approx 1-2$ keV, in which case the
atmospheric microphysics is simplified.  The continuum opacity,
$\kappa_{\rm cont} = \kappa_{\rm abs} + \kappa_{\rm Th}$, where
$\kappa_{\rm abs}$ is the absorption opacity (mostly free-free) and
$\kappa_{\rm Th}$ is the Thomson scattering opacity, is dominated by
electron scattering.  Bound-free transitions are suppressed since most
atoms are fully ionized (Pavlov, Shibanov \& Zavlin 1991; London et
al. 1986).
We do not consider the global problem of the thermal spectra from a
bursting NS (for a review see Lewin, Van Paradijs \& Taam 1993 and
references therein), but rather the simplified problem of line
transport in the Fe layer.  The spectra from a bursting neutron star,
whose opacity is scattering dominated, is expected to be non-Planckian
(Zeldovich \& Shakura 1969; Castor 1974; Madej 1974; van Paradijs
1982).  For our purposes, we take the color correction given by Madej,
Joss \& Rozanska (2004) (also see Madej 1991) to relate the observed
color temperature to the effective temperature.  The effect of the
color correction is nearly offset by the measured redshift correction.
Hence the observed color temperature {\it at infinity} is roughly the
effective temperature {\it at the surface}.  For the sake of
simplicity, we approximate the continuum flux with a blackbody at the
effective temperature.  However, we will discuss the effect of the
spectral shape of the continuum flux on our results in
\S~\ref{sec:meat}.

We assume that the continuum is formed below the line forming region
and is freely streaming. The equation of radiative transfer is then
(Mihalas 1978)
\begin{eqnarray}\label{eq:radiative_transfer}
\mu \frac {\partial I_{\nu}} {\partial y} &=& -\left(\kappa_{\rm abs}
+ 
\kappa_{\rm Th} + \psi_l \phi_{\nu}\right) I_{\nu} + \kappa_{\rm abs}
B_{\nu} +\kappa_{\rm Th} J_{\nu}  \nonumber\\&& + \epsilon \psi_l \phi_{\nu} B_{\nu} +
\left(1 - \epsilon\right)\psi_l \int R(\nu, \nu') J_{\nu'} d\nu',
\end{eqnarray}
where $I_{\nu}$ is the frequency and angle dependent specific
intensity, $J_{\nu} = (1/2) \int I_{\nu} d\mu$ is the angle-average
mean intensity, $B_{\nu}$ is the Planck function, $\psi_l$ is the line
opacity properly normalized, $\epsilon = \kappa_{\rm abs}/\kappa_{\rm
  tot}$ is the absorbed fraction, where $\kappa_{\rm tot} =
\kappa_{\rm cont} + \psi_l$, $\phi$ is the line profile distribution
and is determined by the physics of \S~\ref{sec:Stark}, and $R(\nu,
\nu')$ is the redistribution function.  The coordinate $y$ measures
the column from the base of the scattering layer. 

\subsection{Line Broadening Physics}\label{sec:Stark}

We begin by discussing the line broadening physics.  The number
density of fully ionized H at the photosphere is
\begin{equation}
n_p \approx 2 \times 10^{23} \left(\frac{1\,{\rm keV}}{k_BT_g}\right)
\left(\frac g {3\times 10^{14}\,{\rm cm\,s}^{-2}}\right)\,{\rm cm}^{-3},
\end{equation} 
where $g$ is the surface gravity. It is sufficiently dense so that the
intrinsic broadening of the hydrogenic Fe H$\alpha$ line is dominated
by the linear Stark effect (Paerels 1997).  The linear Stark effect is
only relevant for hydrogenic atoms and breaks down for multi-electron
atoms where the degeneracies of opposite parity states are broken
(Bethe \& Salpeter 1957).  The distribution of the electric
microfields is given by the Holtsmark distribution (Mihalas 1978; also
see Potekhin, Chabrier, \& Gilles 2002) and the characteristic scale
is given from interactions with neighboring protons, $E = e/r_0^2$,
where $r_0 = (4\pi n_p/3)^{-1/3} = 5\times 10^{-9}\rho^{-1/3}\, {\rm
cm}$ is the mean ion spacing and $\rho = n_p m_p$ is the
density. Hence the scale of Stark broadening is
\begin{eqnarray}\label{eq:StarkScale}
\Delta E_{\rm Stark} &=& h\Delta\nu_{\rm Stark} \approx \frac {\hbar^2
  n^2} {m_e Z r_0^2}\nonumber\\ & \approx & 1.5 \,Z_{26}^{-1}
  \left(\frac {n_p} {10^{23}\,{\rm cm}^{-3}}\right)^{2/3} \left(\frac
  n 3\right)^2 \,{\rm eV},
\end{eqnarray}
for the H$\alpha$ line, $0.6\,{\rm eV}$ for the Ly$\alpha$ line, and
$2.7\,{\rm eV}$ for the P$\alpha$ line, where $Z_{26} = Z/26$.
Note that we have chosen the upper state because of the strong
dependence on $n$.  The scale of thermal Doppler broadening is 
\begin{eqnarray}
\Delta
E_{\rm th} &=& h\Delta\nu_{\rm th} = \left(\frac{2k_BT_g}{Am_pc^2}\right)^{1/2}E_{\rm
tran}\nonumber\\
&\approx& 0.25 \left(\frac{k_BT_g}{1\,{\rm keV}}\right)^{1/2} \left(\frac A {56}\right)^{-1/2}
\left(\frac{E_{\rm
tran}}{1.276\,{\rm keV}}\right)\,{\rm eV},
\end{eqnarray} 
where $E_{\rm tran}$ is the line
energy, scaled to the H$\alpha$ transition and $A_{56} = A/56$.  For
Ly$\alpha$ and P$\alpha$, $\Delta E_{\rm th} \approx 1.4$ eV
and $\Delta E_{\rm th} \approx 0.087$ eV respectively.  Stark
broadening dominates the H$\alpha$ line and P$\alpha$, but
thermal Doppler broadening dominates the Ly$\alpha$ line.  The
relative strength of Stark and thermal Doppler broadening scales like
\begin{eqnarray}
\lambda \equiv \frac {\Delta \nu_{\rm th}}{\Delta\nu_{\rm Stark}}
\approx 11 \,Z_{26}^{3} A_{26}^{-1/2} \left(\frac {k_BT_g} {1\,{\rm
keV}}\right)^{1/2}\nonumber\\
\left(\frac {n_p} {10^{23}\,{\rm
cm}^{-3}}\right)^{-2/3} \frac{\left| n^{-2} - n'^{-2}\right|}{n^2},
\end{eqnarray}
where $n$ is the upper state and $n'$ is the lower state.  Note the
dependence on $n$ and $n'$.  Depending on the values of $n$ and $n'$,
the relative strengths of the two broadening mechanisms can vary
tremendously.  For instance for the H$\alpha$ transition, $\lambda
\approx 0.17$, while for the Ly$\alpha$ and P$\alpha$
transitions $\lambda \approx 2$ and $\approx 0.033$ respectively.

The timescale of changes in electric microfields is $\tau_{\rm El} =
r_0/v_{\rm th}$, where $v_{\rm th} = \sqrt{2k_BT_g/m_p}$ is the thermal
speed of the perturbing protons.  At photospheric temperatures and
densities, this gives $\tau_{\rm El} \approx 7 \times 10^{-16}\,{\rm
s}\,(k_BT_g/1\,{\rm keV})^{-1/2}(n_p/10^{23}\,{\rm cm}^{-3})^{-1/3}$,
much smaller than the $n=3\to 2$ radiative deexcitation
timescale of $3.1 \times 10^{-14}\,{\rm s}$. Since the electric field
changes during the time in which the atom absorbs and re-emits a
photon, the energy of the two H$\alpha$ photons are uncorrelated. With
this in mind, we can assume {\it complete redistribution}, greatly
simplifying the line transport calculation.

\begin{figure}
  \plotone{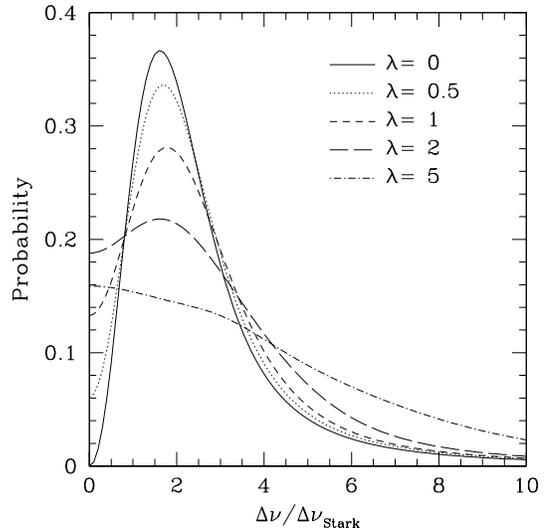}
  \caption{Distribution of frequency shifts due to a combination of
  thermal Doppler (Voigt) and Stark shifts (Holtsmark) at various
  relative strengths of each.  The x-axis is given in units of
  $\Delta\nu_{\rm Stark}$ and the relative strengths of each
  distribution is given by $\lambda = \Delta\nu_{\rm
  th}/\Delta\nu_{\rm Stark}$.  The limits of $\lambda=0$ and
  $\lambda \gg 1$ gives the Holtsmark distribution and the Voigt
  profile respectively.}
  \label{fig:distribution}
\end{figure}

The final broadening profile is a convolution of Stark and thermal
Doppler broadening.  In Figure \ref{fig:distribution}, we show these
profiles for various values of $\lambda$. Asymptotically, the
distribution approaches the power law $\varpropto
\left(\Delta\nu/\Delta\nu_{\rm Stark}\right)^{-2.5}$, expected for the
Holtsmark distribution (Mihalas 1978).  For $\lambda \gg 1$,
we recover the thermal Doppler profile. For $\lambda \ll 1$,
we approach the Holtsmark distribution.  Note that for small values of
$\lambda$, there is a severe deficit at zero shift because this shift
corresponds to perfect symmetry in the proton distribution.  In fact
for $\lambda=0$, the probability of a zero shift goes precisely to
zero.  The effect of thermal Doppler broadening is to fill in the
distribution near zero (see Figure \ref{fig:distribution}).  The
deficit near zero shift yields an emission-like feature there, which
must be interpreted as reduced scattering. This is similar to the
``ghost of Ly$\alpha$'' where reduced scattering in the line
contributes to an emission-like feature in the spectra of broad
absorption line (BAL) quasars (Arav 1996).

\subsection{NLTE Effects}

We now show that the atomic level populations in the photospheres of
bursting neutron stars are not expected to be in LTE (London et
al. 1986, BCP). For simplicity we only consider a two-level system.
The rate equations for the density of upper ($n_u$) and lower ($n_l$)
states are then,
\begin{equation}\label{eq:two-level}
0 = n_l\left[C_{lu} + B_{lu} J(E_{lu})\right] - n_u\left[C_{ul} +
A_{ul} + B_{ul} J(E_{lu})\right],
\end{equation} 
where $C_{lu, ul}$ are the collisional rates from the lower to upper
state and vice versa, $A$ and $B$ are the Einstein coefficients, $E_{lu}$
is the energy of the transition and $J$ is the mean radiation
intensity at that energy. We take the dipole approximation from
Jefferies (1962) for optically allowed transitions for the collisional
rates:
\begin{eqnarray}
C_{lu} = 5.5 \times 10^{12} \left(\frac {E_{lu}}
{k_BT_g}\right)^{-1.68}\exp\left(-\frac {E_{lu}} {k_BT_g}\right)
\nonumber\\\left(\frac {k_BT_g} {1\,{\rm keV}}\right)^{-1.5} \left(\frac {n_e}
{10^{23}\,{\rm cm}^{-3}}\right) f_{lu}\ {\rm s}^{-1},
\end{eqnarray}
where $f_{lu}$ is the oscillator strength for the transition.  Putting
this all together, we have $C_{lu} \approx 2.2 \times 10^9\ {\rm
s}^{-1}$ for the $n=1 \to 2$ transition and $1.3 \times
10^{12}\ {\rm s}^{-1}$ for the $n = 2\to 3$ transition at
$k_BT_g = 1.8\,{\rm keV}$.  If we compare this to the radiative
excitation/deexcitation rates in the same conditions, we find that
$B_{lu} J(E_{lu}) = 7 \times 10^{13}\,{\rm s}^{-1}$ for the $n=1
\to 2$ transition and $R_{lu} = 7 \times 10^{12}\,{\rm
s}^{-1}$, easily exceeding the collisional rates for the $n=2
\to 3$ transition at a radiation temperature of $k_BT_{\rm
eff} = 1.8\,{\rm keV}$.

Having radiative rates exceed collisional rates is not the only
prerequisite to break LTE.  If the radiation field were isotropic and
Planckian, we would recover the Boltzmann distribution for the level
populations where the relevant temperature setting the level
populations is that of the radiation.  Hence to break LTE, we require
the radiation field to be either non-Planckian, anisotropic or both.
This is indeed the case above the photosphere where the radiation
field is free-streaming and the source function deviates from
Planckian due to the nature of scattering dominated atmospheres
(Mihalas 1978).  Therefore, to successfully model the
spectral line formation, we must solve the equations of radiative transfer and
statistical equilibrium simultaneously.

\section{Line Formation Calculations}\label{sec:line_formation_calculations}

We rewrite equation (\ref{eq:radiative_transfer}) by defining the flux
$F_{\nu} = (1/2) \int \mu I_{\nu} d\mu$ and $K_{\nu} = (1/2) \int
\mu^2 I_{\nu} d\mu$.  Integrating equation
(\ref{eq:radiative_transfer}) over $d\mu$ and $\mu d\mu$ we find
(Mihalas 1978),
\begin{eqnarray}
\frac {\partial F_{\nu}} {\partial \tau_{\nu}} &=& -J_{\nu} + \left(1 +
\beta_{\nu}\right)^{-1}\left[\left(1 - \rho +
\epsilon\beta_{\nu}\right) B_{\nu} + \rho J_{\nu}\right.\nonumber\\ &&\left.+ \left(1 -
\epsilon\right)\beta_{\nu} \int R(\nu, \nu') J_{\nu'} d\nu'\right], \\
\frac {\partial K_{\nu}} {\partial \tau_{\nu}} &=& -F_{\nu},
\label{eq:radiative_transfer_split}
\end{eqnarray}
where $d\tau_{\nu} = \left(\kappa_{\rm abs} + \kappa_{\rm Th} + \psi_l
  \phi_{\nu}\right) dy$, $\rho = \kappa_{\rm Th}/\left(\kappa_{\rm
  abs} + \kappa_{\rm Th}\right)$ and $\beta_{\nu} =
  \psi_l\phi_{\nu}/\left(\kappa_{\rm abs} + \kappa_{\rm Th}\right)$.
  Note that $\tau_{\nu}$ is defined from the base of the scattering
  layer.  If we define the variable Eddington factor as $f_{\nu} =
  K_{\nu}/J_{\nu}$, we can write equation
  (\ref{eq:radiative_transfer_split}) as
\begin{eqnarray}
\frac {\partial^2 f_{\nu} J_{\nu}} {\partial \tau_{\nu}^2} = J_{\nu} -
\left(1 + \beta_{\nu}\right)^{-1}\left[\left(1 - \rho +
  \epsilon\beta_{\nu}\right) B_{\nu} + \rho J_{\nu}\right.\nonumber\\\left. + \left(1 -
  \epsilon\right)\beta_{\nu} \int R(\nu, \nu') J_{\nu'} d\nu'\right].
\label{eq:radiative_transfer_var}
\end{eqnarray}
Equation (\ref{eq:radiative_transfer_var}) is a generalization of
Harrington's (1973) equation including the use of variable Eddington
factors.

We solve equation (\ref{eq:radiative_transfer_var}) with a standard
methodology (Mihalas 1978).  Following the nomenclature of Mihalas
(1978) we define a symmetric and antisymmetric average for the
intensity
\begin{eqnarray}
u_{\nu}(\tau_{\nu}, |\mu|) = \frac 1 2 \left[I_{\nu}(\tau_{\nu}, |\mu|) +
I_{\nu}(\tau_{\nu}, -|\mu|)\right] \\ 
v_{\nu}(\tau_{\nu}, |\mu|) = \frac 1 2 \left[I_{\nu}(\tau_{\nu}, |\mu|) -
I_{\nu}(\tau_{\nu}, -|\mu|)\right],
\end{eqnarray} 
which combined with the equation of radiative transfer
(eq.[\ref{eq:radiative_transfer}]) yields
\begin{eqnarray}\label{eq:uv}
{|\mu|} \frac {\partial v_{\nu}} {\partial \tau_{\nu}} &=&  -u_{\nu} + S_{\nu},\\
|\mu|\frac {\partial u_{\nu}} {\partial \tau_{\nu}} &=& -v_{\nu},
\end{eqnarray}
where 
\begin{eqnarray}
S_{\nu} = \left(1 + \beta_{\nu}\right)^{-1}\left[\left(1 - \rho +
\epsilon\beta_{\nu}\right) B_{\nu} + \rho J_{\nu} + \right.\nonumber\\
\left.\left(1 - \epsilon\right)\beta_{\nu} \int R(\nu, \nu') J_{\nu'}
d\nu'\right]
\end{eqnarray} 
is the source function.  To solve this system of equations, we take an
initial guess for the variable Eddington factor, $f_{\nu}$, and assume
an initial input $I_{\nu}$ at the bottom of the scattering layer that
is isotropic in the outward half space ($\mu \geq 0$). We then solve
for $J_{\nu}$ in equation (\ref{eq:radiative_transfer_var}). We
compute $S_{\nu}$ given $J_{\nu}$ and solve for $u_{\nu}$ and
$v_{\nu}$ in equation (\ref{eq:uv}). With $u_{\nu}$ and $v_{\nu}$ in
hand, we compute, $I_{\nu}$, $F_{\nu}$, and $K_{\nu}$, compute a new
$f_{\nu}$ and iterate to convergence.

Since the atmosphere is in NLTE and scattering dominated, we must
simultaneously solve the equations of radiative transfer
(eq.[\ref{eq:radiative_transfer}]) and statistical equilibrium
(eq.[\ref{eq:two-level}]).  For simplicity, we assume a two-level atom
and calculate the level populations between the two states (in the
case of H$\alpha$, the two states are the n=2 and n=3 states; for
Ly$\alpha$, n=1 and n=2; for P$\alpha$, n=3 and n=4) fixing the
sum of the two states.  We iteratively solve the radiative transfer
equation (eq.[\ref{eq:radiative_transfer}]) and use the computed mean
intensities to solve for statistical equilibrium
(eq.[\ref{eq:two-level}]); i.e. $\Lambda$ iteration.  We define
convergence when the mean intensity varied by less than $\delta J/J <
10^{-3}$, which was typically achieved in 3-5 iterations.

\subsection{Line Profile and Equivalent Width}\label{sec:meat}

We solve the thin scattering layer illustrated by Figure
\ref{fig:model} with the above technique.  However, before we
continue, we review the analytic result of a specific case of
scattering dominated atmosphere known as Schuster's law (Schuster
1905). Assuming that line opacity is dominant ($\beta_{\nu}\gg 1$),
the absorbed fraction is small ($\epsilon \to 0$), and the scattering
is coherent ($R(\nu',\nu)=\delta_{\nu',\nu}$), equation
(\ref{eq:radiative_transfer_var}) becomes
\begin{equation}
\frac {\partial^2 f_{\nu} J_{\nu}} {\partial \tau_{\nu}^2} = 0.
\end{equation}
Solving this equation, we find $f_{\nu}J_{\nu} = A_{\nu} \tau_{\nu} +
B_{\nu}$, where $A_{\nu}$ and $B_{\nu}$ are constants.  From equation
(\ref{eq:radiative_transfer_split}), we also find $F_{\nu} =
f_{\nu}A_{\nu}$.

We now make some further approximations to determine $A_{\nu}$ and
$B_{\nu}$.  We assume the Eddington approximation everywhere so that
$f_{\nu} = 1/3$ (Rybicki \& Lightman 1979).  We also assume the two
stream approximation, which breaks the radiation field into one
outgoing stream ($I_{\nu}^+= I_{\nu}(\tau_{\nu}, |\mu|)$) and one
incoming stream ($I_{\nu}^+= I_{\nu}(\tau_{\nu}, -|\mu|)$).
Consistency with the Eddington approximation demands
$|\mu|=1/\sqrt{3}$ (Rybicki \& Lightman 1979).  The boundary
conditions are that the incoming flux at the surface is zero,
$I_{\nu}(\tau_{\nu} = \tau_{\nu}^{\rm tot}, \mu = -1/\sqrt{3}) = 0$,
where $\tau_{\nu}^{\rm tot}$ is the total optical depth of the
scattering layer at frequency $\nu$ and that below the line forming
region, the incident intensity is fixed at some value typical for Type
I bursts ($I_{\nu}(\tau_{\nu} = 0, \mu = 1/\sqrt{3}) = I_{\rm inc}$).
Per frequency the optical depth of the slab is given by
$\tau_{\nu}^{\rm tot} = \tau_{0}^{\rm tot}\phi(\nu)$, where
$\tau_{0}^{\rm tot}$ is the normalization of the optical depth and is
given by
\begin{eqnarray}
\tau_{0}^{\rm tot} &=& N_{\rm Fe, n=2} \sigma_{0,23} \nonumber\\
&\approx&6.1 \left(\frac
{N_{\rm Fe, n=2}}{10^{17}\,{\rm cm}^{-2}}\right) \left(\frac
{n_p}{10^{23}\,{\rm cm}^{-3}}\right)\left(\frac{Z}{26}\right)^{-1},
\end{eqnarray}
where $\sigma_{0,23} = \pi e^2 f_{23}/(m_e c \Delta \nu_{\rm Stark})$
is the cross section at line center. In the two-stream approximation,
our definitions for the mean intensity and the flux become
\begin{eqnarray}
J_{\nu} &=& \frac 1 2 \left(I_{\nu}^+ + I_{\nu}^-\right) \\
F_{\nu} &=& \frac 1 {2\sqrt{3}} \left(I_{\nu}^+ - I_{\nu}^-\right)
\end{eqnarray}
and hence solving for $A_{\nu}$ and $B_{\nu}$, we find
\begin{equation}\label{eq:schuster}
I_{\rm trans} = I_{\rm inc} \frac {2} {\sqrt{3}\tau_{\nu}^{\rm
tot} + 2},
\end{equation}
where $I_{\rm trans} = I_{\nu}(\tau_{\nu} = \tau_{\nu}^{\rm tot}, \mu
= 1/\sqrt{3})$ is the transmitted intensity. Equation
(\ref{eq:schuster}) is known as Schuster's law.  Note that at large
$\tau_{\nu}^{\rm tot}$, the specific intensity scales with
$1/\tau_{\nu}^{\rm tot}$ unlike the absorption dominated case where
$I_{\nu} \to B_{\nu}$ at large $\tau_{\nu}^{\rm tot}$.  This
difference in scaling with optical depth between a scattering
dominated atmosphere and an absorption dominated atmosphere is the key
to understanding the nature of the Fe H$\alpha$ line.

\begin{figure}
  \plotone{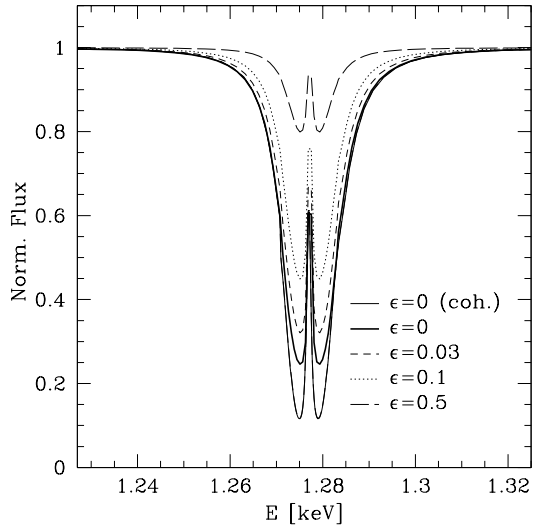}
  \caption{Hydrogenic Fe H$\alpha$ line profiles of various values of
  $\epsilon$ in the line at the NS surface. For the background we take
  $\tau_{0}^{\rm tot} = 100$, $k_BT_{\rm eff} = 1.8\,{\rm keV}$ and
  $n_p = 10^{23}\,{\rm cm}^{-3}$. The numerical calculation of two
  stream approximation and exact analytic result
  (eq[\ref{eq:schuster}]) are indistinguishable and are represented by
  the thin-solid line. The full numerical results properly angle
  averaged for complete redistribution are represented by the other
  curves for various values of the absorption ratio $\epsilon$.  The
  difference between the thin-solid line and the thick-solid line are
  due to the departures of the two-stream approximation compared to
  the full angle-averaged solution and coherent scattering compared to
  complete redistribution.  The difference between the other curves
  are due to different values of $\epsilon$.}
  \label{fig:coherent}
\end{figure}

We compare the flux from our exact numerical calculations with
Schuster's law (eq.[\ref{eq:schuster}]) in Figure \ref{fig:coherent}.
For simplicity of the discussion, we take a simple model of a
hydrogenic atom whose energy levels are defined by $E_n = -13.6\,Z^2
n^{-2}\,{\rm eV}$, which gives the energy of the Fe H$\alpha$
transition of $1.2769\,{\rm keV}$.  We will consider the fully
relativistic atom with the complications of fine structure splitting
later. We take $\tau_{0}^{\rm tot} = 100$, $k_BT_{\rm eff} = 1.8\,{\rm
keV}$ and $n_p = 10^{23}\,{\rm cm}^{-3}$ for these models.  Our
numerical calculation for coherent scattering in the two-stream
approximation and Schuster's law are indistinguishable and are both
represented by the thin-solid line.  This confirms the accuracy of our
numerical treatment with the exact analytic result.  The other
calculations represent the full numerical solutions with {\it complete
redistribution} with the variable Eddington factor, $f_{\nu}$,
computed self-consistently.  These curves are computed with 10 angular
points, 100 frequency points and 100 depth points at various values of
the absorption coefficient $\epsilon$.  The difference between the
thick-solid line and the thin-solid line is due to differences
between coherent scattering and complete redistribution and the nature
of the two-stream approximation which fixes $f_{\nu} = 1/3$ at every
point compared to the full self-consistent calculation.  Note the
emission-like feature at line center, which is due to the deficit near
zero shift discussed in \S~\ref{sec:Stark}.

Now that we have established the effectiveness of our numerical code,
we discuss the nature of the hydrogenic Fe Ly$\alpha$, H$\alpha$ and
P$\alpha$ lines.  Since we have limited our study to two-level
atomic systems, the relevant transitions starting from the 1s ground
state are 1s $\rightleftarrows$ 2p, 2p $\rightleftarrows$ 3d and 3d
$\rightleftarrows$ 4f.  The fine structure splitting of these
transitions significantly complicates the line structure\footnote{We
thank Rashid Sunyaev for raising this important issue.}.  Due to fine
structure splitting the energy levels of an hydrogenic atom are
\begin{equation}
E_{n,j} = -13.6 \frac {Z^2} {n^2}\left[1 + \frac {(\alpha Z)^2}{n}
\left(\frac 1 k - \frac 3 {4n}\right)\right]\,{\rm eV},
\end{equation}
where $j$ is the total angular momentum quantum number,
\begin{equation}
k = \left\{\begin{array}{ll} 
l & \textrm{if } j = l - \frac 1 2\\
l+1 & \textrm{if } j = l + \frac 1 2,
\end{array}\right.
\end{equation}
$l$ is the orbital quantum number, and $\alpha$ is the fine structure
constant. The Ly$\alpha$ 1s $\rightleftarrows$ 2p transition is fine
structure split into two lines with relative line intensities of 2:1
with energies of 6.9728 keV and 6.9521 keV, a difference of $20.7$ eV.
The H$\alpha$ 2p $\rightleftarrows$ 3d transition is split into three
lines with relative line intensities of 9:1:5 and energies of 1.2811
keV, 1.2790 keV, and 1.2997 keV.  Taking the energy of the strongest
line to be the centroid, the splitting of the levels are 0, -2.1, 18.6
eV (Bethe \& Salpeter 1957).  The size of the splitting is comparable
to the EW and the FWHM of the observed line.  Finally the
P$\alpha$ 3d $\rightleftarrows$ 4f transition is split into
three lines with intensities 20:1:14 and energy differences of 
0, -0.43, 1.61 eV.  These splittings are smaller
than the effect of Stark broadening, 
so that their effect on the line profile is small.

\begin{figure}
  \plotone{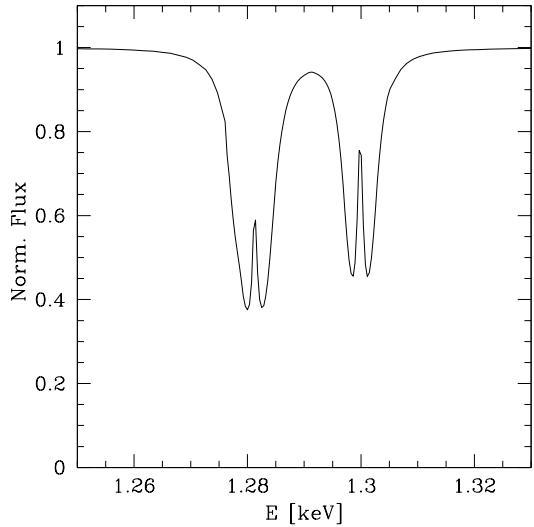}
  \caption{Hydrogenic Fe H$\alpha$ line profiles with fine structure
  splitting. We take the same background as Figure
  \ref{fig:coherent} ($k_BT_{\rm eff} = 1.8\,{\rm keV}$ and $n_p =
  10^{23}\,{\rm cm}^{-3}$).  We also simultaneously solve statistical
  equilibrium assuming an initial two-level system with everything in
  the n=2 state with an optical depth normalization of $\tau_{0}^{\rm
  tot} = 100$.  With these parameters, statistical equilibrium gives
  $\epsilon \approx 0.033$.}
  \label{fig:fine_structure}
\end{figure}

We illustrate the line profile with fine structure splitting in Figure
\ref{fig:fine_structure}.  We also turn on statistical equilibrium so
that from an initial state where the initial population of the
two-level atom was assumed to be in the n=2 state (with an optical
depth normalization of $\tau_{0}^{\rm tot} = 100$), we simultaneously
solve radiative transfer and statistical equilibrium.  Since the
opacity is dominated by line itself, the absorbed fraction is
determined from the ratio of the collisional rate to the total
transition rates ($\epsilon \equiv \kappa_{\rm abs}/\kappa_{\rm tot}
\approx C_{lu}/(R_{lu} + C_{lu})$).  In statistical equilibrium,
$\epsilon$ relaxes to a small value of $0.033$, typical of a
scattering dominated atmosphere.  This line profile represents the
complete solution to the spectra from the resonant scattering layer.
Note that the triplet structure of the fine structure lines is reduced
to a doublet because Stark broadening is so large.

\begin{figure}
\plotone{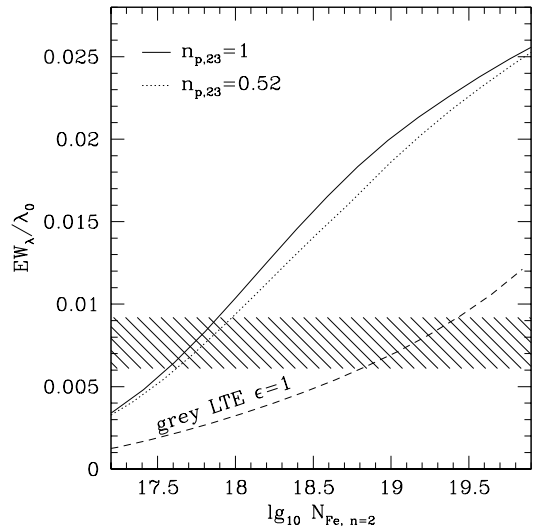}
  \caption{Fractional Equivalent width as a function of the Fe column
  in the $n=2$ state for two different proton densities of $n_p =
  5.2\times 10^{22}\,{\rm cm}^{-3}$ (dotted line) and $n_p =
  10^{23}\,{\rm cm}^{-3}$ (solid line) at $k_BT_{\rm eff} = 1.8\,{\rm
  keV}$. We show the observed EW as a banded region and the
  intersection of these lines with the banded region indicates the
  $n=2$ Fe column needed. We also present an extremal LTE calculation
  (dashed line) where $A_0\approx 0.2$ (BCP) for comparison.}
  \label{fig:ew}
\end{figure}

The line's equivalent width (EW) is the proportional deficit and is
defined as
\begin{equation}\label{eq:EW_defined}
EW \equiv \int \left(1 - \frac {F_{\nu}^{\rm line}} {F_{\nu}^{\rm
cont.}}\right) d\nu.
\end{equation}
Though rotation will modify the line profile, it conserves equivalent
width (Gray 1992) to first order in $\Omega R/c$, where $\Omega$ is
the angular rotation rate.  In Figure \ref{fig:ew}, we show the
fractional equivalent width, defined as the equivalent width in
wavelength units over the line center wavelength, as a function of the
Fe column in the $n=2$ state for $n_p = 5.2\times 10^{22}\,{\rm
cm}^{-3}$ (dotted line) and $n_p = 10^{23}\,{\rm cm}^{-3}$ (solid
line), corresponding to the Coulomb stopping density for radial infall
and the $\tau \approx 1$ surface (approximating a solar metallicity
atmosphere) for a radiation temperature of $k_BT_{\rm eff} = 1.8\,{\rm
keV}$.  We also show the observed EW (CPM).  The scattering dominated
calculations imply $N_{\rm Fe, n=2} \approx 5 \times 10^{17}\,{\rm
cm}^{-2}$. We also plot an extremal grey atmosphere model in perfect
LTE ($\epsilon = 1$) for comparison.  The LTE calculation requires a
much larger $N_{\rm Fe, n=2}$ and highlights how poor the LTE
calculation fares compared to the full resonant calculation.

\begin{figure}
  \plotone{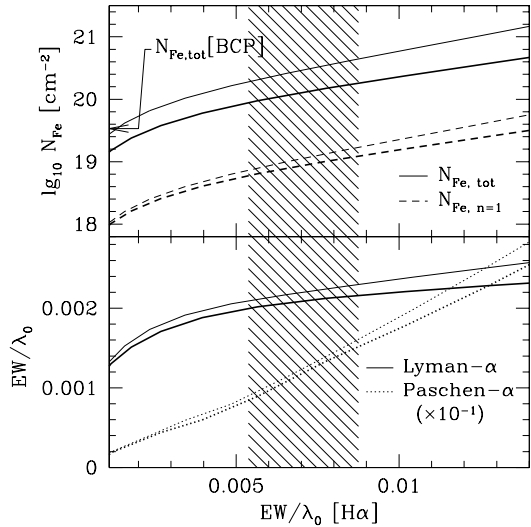}
  \caption{$N_{\rm Fe,n=1}$, $N_{\rm Fe}$, EW$_{\rm Ly\alpha}$, and
  EW$_{\rm P\alpha}$ as a function of total EW$_{\rm H\alpha}$. for
  parameters $n_p = 5.2 \times 10^{22}\,{\rm cm}^{-3}$ (thin lines),
  $n_p = 10^{23}\,{\rm cm}^{-3}$ (thick lines) and $k_BT_{\rm eff} =
  1.8\,{\rm keV}$. The banded region represents the observed EW and
  the intersection of these lines with the banded region indicates the
  Fe column (upper plot) needed in the hydrogenic ground state (dashed
  lines) and total Fe (solid lines) and the predicted EW in the
  Ly$\alpha$ line (solid lines; lower plot) and P$\alpha$ line
  (dotted lines; lower plot).  The P$\alpha$ values have been
  divided by 10 so that they may fit on the same plot.}
  \label{fig:spectral}
\end{figure}

We calculate a grid of models for the Ly$\alpha$, H$\alpha$ and P$\alpha$
transitions to relate the equivalent widths and level populations in each 
transition to one another.  We estimate the total $N_{\rm Fe}$ to
that in the n=1 state via Saha equilibrium. 
Since the radiation
field drives ionization, we expect our results from Saha equilibrium
will give a {\it lower} bound on $N_{\rm Fe}$.  We relate EW$_{\rm
H\alpha}$, $N_{\rm Fe, n=1}$, $N_{\rm Fe}$, EW$_{\rm Ly\alpha}$ and
EW$_{\rm P\alpha}$ in Figure \ref{fig:spectral} at two
different proton densities: $n_p = 5.2 \times 10^{22}\,{\rm cm}^{-3}$
(Fe stopping layer; dotted line) and $n_p = 10^{23}\,{\rm cm}^{-3}$
(solid line).  The larger proton density is indicative of what would
happen in a uniform metallicity atmosphere.  The observed Fe H$\alpha$
EW gives an implied column of $N_{\rm Fe, n=1} \approx 1-2 \times
10^{19}\,{\rm cm}^{-2}$ if all the Fe is concentrated in the Fe
stopping layer.  Saha equilibrium then implies a total Fe column of
$N_{\rm Fe}=3-6 \times 10^{20}\,{\rm cm}^{-2}$ at the Fe stopping
layer, a factor of $\approx 10$ higher than the solar metallicity
accretion/spallation scenario.  For the larger proton density ($n_p =
10^{23}\,{\rm cm}^{-3}$), the implied Fe column is $\approx
10^{20}\,{\rm cm}^{-2}$, a factor of two larger than that implied by a
uniform solar metallicity photosphere.  Since ionization balance is
likely given by the radiation field (London et al. 1986), the {\it
true} $N_{\rm Fe}$ is likely even higher for both these scenarios.  We
also estimate that our implied $N_{\rm Fe, n=1}$ gives an EW$_{\rm
Ly\alpha} \approx 15-20\,{\rm eV}$ (redshifted 11-15 eV).  Though this
is larger than EW$_{\rm H\alpha}$, rotational broadening has a large
impact on its detectability (see \S~\ref{sec:rotation}).  For the
P$\alpha$ line, the EW$_{\rm P\alpha} \approx 4-7$ eV
(redshifted 3-5 eV) for the observed EW$_{\rm H\alpha}$.

\begin{figure}
\plotone{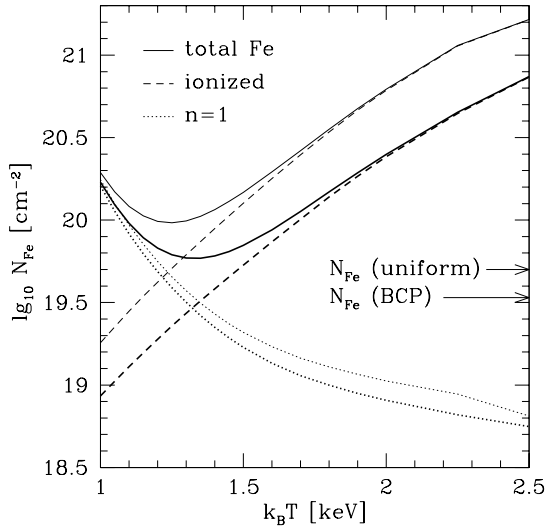}
  \caption{The Fe column required to generate the observed EW in the
  Fe H$\alpha$ line as a function of $k_BT_{\rm eff}$ at the Fe
  stopping column of $y_{\rm stop} \approx 1\,{\rm g\,cm}^{-2}$ (thin
  lines) and $y_{\rm stop} = 2\,{\rm g\, cm}^{-2}$ (thick lines),
  which is more representative of a uniform atmosphere.  The various
  Fe columns represented are the hydrogenic Fe column (dotted-line),
  fully ionized Fe column (dashed-line) and total Fe column
  (solid-line). We find that $N_{\rm Fe}$ must at least be a factor of
  3-4 larger than that predicted from the accretion/spallation
  scenario of $3.4\times 10^{19}\,{\rm cm}^{-2}$ (BCP), but is within
  a factor of 2 of a uniform solar metallicity atmosphere of $5 \times
  10^{19}\,{\rm cm}^{-2}$.}
  \label{fig:kTVsNFe}
\end{figure}

We now take $k_BT_{\rm eff}$ as a free parameter to understand how our
results depend on the color correction and spectra. We plot the
implied n=1 and total $N_{\rm Fe}$ as a function of $k_BT_{\rm eff}$
at the observed EW in Figure \ref{fig:kTVsNFe}.  Since we are
interested in Fe at the stopping column (measured from the top of the
atmosphere) of $y_{\rm stop} \approx 1\,{\rm g\,cm}^{-2}$ (thin
lines), the local proton density, which is important for Stark
broadening, is then $n_p \approx 9.4 \times 10^{22} \left(k_BT_{\rm
    eff}/1\,{\rm keV}\right)^{-1}\,{\rm cm}^{-3}$.  The implied Fe
column is minimized at $k_BT_{\rm eff}=1.3\,{\rm keV}$, but it is 3-4
times larger than expected from our solar metallicity
accretion/spallation scenario.  We also plot a similar Fe column at
$y_{\rm stop} = 2\,{\rm g\,cm}^{-2}$ (measured from the top of the
atmosphere), which is more representative of a uniform atmosphere.
From Figure \ref{fig:spectral} and \ref{fig:kTVsNFe}, we find that our
the implied Fe column for a solar metallicity atmosphere is within a
factor of 3 for the entire range of burst temperatures of 1-2 keV.  We
discuss these points further in \S~\ref{sec:conclusion}.

\subsection{Rotational Broadening}\label{sec:rotation}

\begin{figure}
\plotone{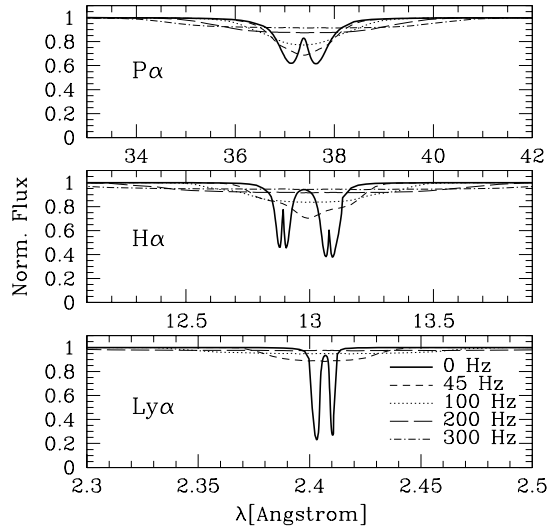}
  \caption{Redshifted line profiles for the Fe P$\alpha$,
  H$\alpha$, and Ly$\alpha$ at various rotation rates. We fixed the Fe
  H$\alpha$ EW to the observed values and assume that the Fe is
  concentrated in the accretion/spallation stopping layer.  We compare
  models (from bottom to top) at $\nu_{\rm spin} \equiv (1+z)^{-1}\Omega/(2\pi)
  = $ 0 Hz (thick-solid lines), 45 Hz (short-dashed lines), 100 Hz
  (dotted lines), 200 Hz (long-dashed lines), and 300 Hz
  (dashed-dotted lines) for $\sin i = 1$.  We do not include a 300 Hz
  Ly$\alpha$ line profile as it is totally rotationally broadened
  away.}
  \label{fig:rotation}
\end{figure}

Rotational broadening heavily modifies the structure of the
lines.  We estimate its effects on the line profile as follows.  Since
the photon occupation number $n \varpropto I_{\nu}/\nu^3$ is invariant
under Lorentz transformations, the effect of rotational broadening on
the emergent NS flux is
\begin{eqnarray}
F_{\nu}^{\rm obs} &=& \frac 1 {2\pi} \int_0^{2\pi} d\phi \int_0^1
d(\cos \theta) \cos\theta \nonumber\\ &&\left(1 - (\Omega R/c)
\sin\theta \cos\phi \sin i\right)^{-3}\nonumber\\ &&I\left(\nu(1 -
(\Omega R/c) \sin\theta \cos\phi \sin i),\right.\nonumber\\
&&\left. \cos\theta(1 + (\Omega R/c) \sin\theta \cos\phi \sin
i)\right) .
\end{eqnarray}
Following various authors (Chandrasekhar 1945), we ignore the effect
of Lorentz boosts on the angle and focus on the effects of the
frequency shift. We are currently calculating the exact general
relativistic calculation elsewhere (Morsink et al. 2005; also see 
\"Ozel \& Psaltis 2003 and Bhattacharyya, Miller \& Lamb 2004). In addition
for narrow lines, $|(dI_{\nu}/d\ln\nu)(\Omega R/c)| \gg 1$, so
we can ignore the effects of the $(1 - \Omega R/c)^3$ in the
denominator, leaving
\begin{eqnarray}
F_{\nu}^{\rm obs} &=& \frac 1 {2\pi} \int_0^{2\pi} d\phi \int_0^1
d(\cos \theta) \cos\theta\,\nonumber\\ &&I\left(\nu\left[1 - \frac
{\Omega R} {c} \sin\theta \cos\phi \sin i\right], \cos\theta\right),
\end{eqnarray}
showing that rotational line broadening goes like $(\Omega R/c) \sin
i$.  We plot the Fe P$\alpha$, H$\alpha$ and Ly$\alpha$ lines for a
1.4\Msun\ NS with R=9. 2km at various rotation rates up to 300 Hz in
Figure \ref{fig:rotation}. To ease comparison with observation, we
plot the lines as a function of wavelength and include the redshift
correction of $z = 0.35$.  We also include the $1+z$ redshift
correction to the rotation rate.  For the highest rotation rates
($>200\,{\rm Hz}$), the Fe H$\alpha$ central line depth would only be
about 5-10\% below the continuum and hence the line would be difficult
to detect at the observed EW.  At lower rotation rates the doublet
nature of the line gives an asymmetric profile with a sharp feature.
For the P$\alpha$ line, the fine structure splitting in the line is
smaller than the Stark effect so it appears as a singlet.

Rotational broadening is more pronounced on the Ly$\alpha$ transition,
which has a FWHM $\approx 130\,{\rm eV}$ at a spin frequency of $45$
Hz.  This is 5 times larger than the rotational FWHM for the H$\alpha$
line due to the larger energy of the Ly$\alpha$ transition compared to
the H$\alpha$ transition.  At the implied EW$_{\rm Ly\alpha}$ matched
to the observed EW$_{\rm H\alpha}$, the line is extremely shallow with
a central line depth of approximately 10\%.  Hence even at this modest
rotation rate, the Ly$\alpha$ line may be difficult to detect.  For
larger rotation rates ($>100$ Hz), Ly$\alpha$ is so distributed that
it may be impossible to detect.

Observations of dips and eclipses on EXO~0748-676 (Parmar
et. al. 1986) have revealed that the orbit of the secondary is nearly
edge-on ($\sin i \approx 1$).  Assuming the rotation axis of the NS is
aligned with the orbit, we favor very low rotation rates ($50-100$ Hz)
for EXO~0748-676 compared to the rotation rates that are typically
observed in other LMXB systems where the rotation rate is typically
$300-600$ Hz (for a review see Strohmayer \& Bildsten 2003).  The
recent observation of a 44.7 Hz burst oscillation (Villarreal \&
Strohmayer 2004) is thus consistent with the observation of the line,
assuming that this oscillation is indicative of the NS rotation rate.
It raises the exciting prospect of fitting the line with an
appropriate spectral calculation to derive the NS radius (Morsink et
al. 2005).  Other accreting NSs may also possess atomic lines, but we
would not expect them to be detected unless they rotate slowly ($<
100\,{\rm Hz}$) or are highly inclined.

\subsection{Radiative Levitation}\label{sec:rad_levitation}

BCP noted that Fe in the atmosphere may be sustained by radiative
levitation similar to what occurs in hot stars (Michaud 1970; Michaud
et al. 1976) and white dwarfs (Vauclair, Vauclair \& Greenstein 1979;
Vennes et al. 1988; Chayer, Fontaine \& Wesemael 1995).  To begin, we
calculate the radiative force on a hydrogenic Fe atom, $F_{\rm rad} =
c^{-1}\int {\sigma_{\nu} F_{\nu}}\ d\nu,$, from reflection of the
incident flux, $F_{\nu}$, where $\sigma_{\nu}$ is the cross section
and is defined as $\sigma_{\nu}=\sigma_{0,12}\phi_{\nu}$,
$\sigma_{0,12}=\pi e^2 f_{12}/(m_e c\Delta\nu)$, $\phi_{\nu}$ is the
convoluted Holtsmark profile, and $f_{12}=0.42$ is the oscillator
strength for the Ly$\alpha$ transition (Bethe \& Salpeter
1957). Assuming that the flux does not change as we integrate across
the line, $\int\sigma_{\nu} F_{\nu}\ d\nu = F_{\nu} \pi e^2
f_{12}/(m_e c)$. For a radiation field that is isotropic in the
outward hemisphere, $F_{\nu} = \pi B_{\nu}(T)$ (Rybicki \& Lightman
1979), $B_{\nu}(T)$ is the Planck function which we approximate with
the Wien tail, we find.
\begin{equation}
\frac {F_{\rm rad}} {A m_p g} \approx  9 \times 10^3 \exp\left(-\frac
{6.9\,{\rm keV}} {k_BT_{\rm eff}}\right) A_{56}^{-1} g_{14}^{-1},
\end{equation}
where $g_{14} = g/10^{14}\,{\rm cm\,s}^{-2}$.  For $k_BT_{\rm eff} =
1\,{\rm keV}$, the radiation force on the Ly$\alpha$ exceeds gravity
by a factor of 9 if all the Fe is hydrogenic.

Since the radiation force exceeds that of gravity, the Fe will not
necessarily sediment. Rather, Fe can be sustained in the atmosphere,
allowing for a potential concentration. As the column of Fe builds,
more and more of the line flux is absorbed which decreases the
radiative force.  This continues until an equilibrium column is
reached where enough of the line flux is absorbed that the Fe feels
the same net force as a background proton, so that both species have
the same scale height (Chayer, Fontaine \& Wesemael 1995).  This is
achieved when the radiative force is equal to the effective force
$F_{\rm rad} = Am_p g_{\rm eff} = \left[A - (Z+1)/2\right]m_pg$, which
takes into account the effects of both gravity and the electric field
for a hydrogen atmosphere and assumes all the Fe is hydrogenic (Vennes
et al. 1988; Chayer, Fontaine \& Wesemael 1995).

We now estimate the equilibrium column sitting above the continuum photosphere.
The radiative force on a scattering layer is
\begin{equation}\label{eq:radiative_force}
F_{\rm rad} = \sum_{\rm lines} \frac {F_{\nu_i, {\rm cont}} EW_i} {c}, 
\end{equation}
where $\nu_i$ is the energy of the ith line, $F_{\nu_i, {\rm cont}}$
is the continuum flux and $EW_i$ is the equivalent width of the ith
line.  For pure coherent scattering, Schuster's law
(eq.[\ref{eq:schuster}]) and the definition of the equivalent width
(eq.[\ref{eq:EW_defined}]) gives
\begin{equation}\label{eq:EW_analytic}
  EW_i = \int \frac {\tau^{\rm tot}_{0, i} \phi_{\nu, i}} {2/\sqrt{3}
  + \tau^{\rm tot}_{0, i} \phi_{\nu, i}} d\nu,
\end{equation}
where $\tau^{\rm tot}_{0, i}$ and $\phi_{\nu, i}$ is the normalization
of the optical depth and line profile distribution of the ith line
respectively.  We find an extremely simple form for equation
(\ref{eq:EW_analytic}) by taking the asymptotic form\footnote{The
standard definition is $\phi_{\nu} \approx 1.5 \left(|\nu -
\nu_0|/\Delta\nu_{\rm Stark}\right)^{-2.5}$ for large $\left(|\nu -
\nu_0|/\Delta\nu_{\rm Stark}\right)$. The numerical prefactor of 0.75
comes from the fact that the line is symmetric and hence the
distribution runs from $-\infty$ to $\infty$ and not from $0$ to
$\infty$} of the Holtsmark distribution $\phi_{\nu} \approx 0.75
\left(|\nu - \nu_0|/\Delta\nu_{\rm Stark}\right)^{-2.5}$.  Assuming
$\tau_0^{\rm tot}$ is large, we break up the integral in equation
(\ref{eq:EW_analytic}) into a regime where $\tau_0^{\rm
tot}\phi_{\nu}$ is large compared to $2/\sqrt{3}$ and a regime where
$\tau_0^{\rm tot}\phi_{\nu}$ is small compared to $2/\sqrt{3}$.  Thus,
we get
\begin{eqnarray}\label{eq:EWscaling}
  EW_i &\approx& 2\tau^{\rm tot}_{0, i}\Delta \nu_{\rm
    Stark}\left[\int_{0}^{\eta_0} \frac {d\eta} {\tau_0^{\rm tot}} +
    \int_{\eta_0}^{\infty} \frac {0.75\eta^{-2.5}} {2/\sqrt{3}}
    d\eta\right],\nonumber\\ &\approx & 2.8\Delta\nu_{\rm
    Stark}\left(\tau_{0, i}^{\rm tot}\right)^{0.4}
\end{eqnarray}
where $\eta = |\nu - \nu_0|/\Delta\nu_{\rm Stark}$ and
$0.75\tau_0^{\rm tot}\eta_0^{-2.5} = 2/\sqrt{3}$.  We extend this
simple form for the EW to handle fine structure splitting if the split
lines are widely separated and therefore distinct.  We sum over each
split line to get
\begin{equation}
EW_i \approx 2.8 \Delta\nu_{\rm Stark}\left(\tau_{0, i}^{\rm tot}\right)^{0.4}
\sum_j s_{i,j}^{0.4},
\end{equation}
where $s_{i, j}$ are the relative strength of each split line for the
ith transition.  For Ly$\alpha$, which is split into two lines with a
ratio of 2:1, we find $EW_{{\rm Ly}\alpha} = 4.2 \Delta\nu_{{\rm
Stark,\, Ly}\alpha}\left(\tau^{\rm tot}_{0, {\rm
Ly}\alpha}\right)^{0.4}$.  Similarly for H$\alpha$ with a line
strength ratio of 9:1:5, $EW_{{\rm H}\alpha} = 5.2\Delta\nu_{{\rm
Stark,\,H}\alpha} \left(\tau^{\rm tot}_{0, {\rm
H}\alpha}\right)^{0.4}$.

Following our previous discussion, we now equate the radiative force
(eq.[\ref{eq:radiative_force}]) to the effective force to find the
equilibrium column. For simplicity, let us first consider only the
Ly$\alpha$ line and a radiation field that is isotropic in the outer
half hemisphere.  We find
\begin{equation}\label{eq:rad_lev_column}
N_{\rm Fe, tot} \approx 5 \times 10^{20} \left(\frac g {3\times
10^{14}\,{\rm cm\,s}^{-2}}\right)^{5/3}\xi^{-2/3}\,{\rm cm}^{-2},
\end{equation}
for $kT \approx 2 \,{\rm keV}$, where $\xi \equiv N_{\rm
Fe,tot}/N_{\rm Fe, n=1}$ is the ratio of total Fe column to ground
state hydrogenic Fe column.  The Ly$\alpha$ equivalent width that is
associated with this Fe column is
\begin{equation}
  EW_{{\rm Ly}\alpha} \approx 400 \left(\frac{g}{3 \times 10^{14}\,{\rm
        cm\,s}^{-2}}\right)^{2/3} \xi^{-2/3}\,{\rm eV}.
\end{equation}

Radiative levitation can easily sustain an atmosphere whose
metallicity can be supersolar.  In Saha equilibrium, the ratio between
fully ionized Fe and the n=1 hydrogenic state is 10-20 and $N_{\rm
Fe,\, tot} = 7-12 \times 10^{19}\,{\rm cm}^{-3}$, larger than a
constant solar metallicity atmosphere $N_{\rm Fe} \approx 5 \times
10^{19}\,{\rm cm}^{-2}$ (BCP). These values are also tantalizingly
close to what would be required to for the measured H$\alpha$ line,
but we note that the radiatively levitated Fe column is sensitively
dependent on the background continuum and surface gravity.  Since
bursting NS atmospheres are expected to be harder than a blackbody as
discussed earlier in \S~\ref{sec:model_atmosphere}, we would expect
these values to be lower limits.  The associated Ly$\alpha$ EW for
these Fe columns in this approximation is 50-90 eV.

Other transitions also contribute to the radiative force on the Fe
column.  We now show that their contributions are small compared to
the Ly$\alpha$ radiative force.  First, we consider the contribution
from the H$\alpha$ transition.  For a Boltzmann distribution between
the n=1 and the n=2 state at $kT = 2 \,{\rm keV}$, the column density
in the n=2 state is $N_{\rm Fe, n=2} = 0.1 N_{\rm Fe, n=1}$.  The
radiative force from the H$\alpha$ transition is $F_{{\rm rad,
H}\alpha} \approx 2 \times 10^{12}\,{\rm g\ cm\ s^{-2}}$ which is a
factor of 5 smaller than the radiative force on the Ly$\alpha$
transition of $F_{{\rm rad, Ly}\alpha} \approx 10^{13}\,{\rm g\ cm\
s^{-2}}$.  The contribution from the H$\alpha$ transition increases
the column by about a third, but we will ignore this contribution in
keeping with the order of magnitude spirit of this calculation.  Other
transitions also contribute, but their radiative forces are weaker as
their fluxes are weaker by the Rayleigh-Jeans factor of $\nu^2$
compared to H$\alpha$.

\section{Discussion and Conclusions}\label{sec:conclusion}

Motivated by the accretion/spallation scenario that we first presented
in BCP, we have calculated the resonant radiative transfer of a thin
scattering layer above the NS photosphere.  While our calculation and
microphysics are geared toward understanding the feature as an Fe
H$\alpha$ line, the basic physics would remain the same for other
atomic transitions.  Namely, if the line is produced near the
continuum photosphere where the densities are relatively high, Stark
broadening is dominant, NLTE effects are prevalent and the line is a
result of resonant scattering.  Applied to Fe H$\alpha$, our
calculations show that the EW of the Fe H$\alpha$ line observed by CPM
from \EXO\ requires an Fe column 3-10 times larger than the
accretion/spallation scenario predicts for accretion of solar
metallicity material (BCP).  We also approximate a uniform solar
metallicity atmosphere by assuming the Fe sits at a larger depth.  In
this case, we find that the required column of Fe is within a factor
of 3 of a solar metallicity atmosphere over a large range of
temperatures.

The fact that the H$\alpha$ line is narrow requires that the rotation
rate of \EXO\ is slow. Larger spin rates, assuming a neutron star
radius of around 10 km, would wash out such a line unless the surface
emission is confined to a narrow region around the spin axis
(Bhattacharyya, Miller \& Lamb 2004).  A slow rotation rate for \EXO\
is consistent with the recently measured 44.7 Hz burst oscillation
(Villarreal \& Strohmayer 2004). With these line profiles in hand and
assuming that the burst oscillation of 44.7 Hz is the rotation rate of
the NS, detailed fits to the data raises the exciting prospect of
fitting for the neutron star radius (Morsink et al. 2005).

Since the fine structure splitting of Fe H$\alpha$ is of order the
size of the rotational broadening, we cannot assume that the line is
narrow compared to the rotational broadening.  Previous work (\"Ozel
\& Psaltis 2003; Bhattacharyya, Miller \& Lamb 2004) has presumed that
the intrinsic line profile does not matter because rotational
broadening is dominant.  While true for NS rotating at 300-600 Hz, it
is not the case for a Fe H$\alpha$ line at 45 Hz for a 10 km NS.
Rotational broadening may be the largest effect, but it is by no means
the dominant effect. Therefore, the intrinsic line profile must be
known to apply the observed lines to the study of NS physics.

We also estimated that the Fe EW$_{\rm Ly\alpha}$ and EW$_{\rm
P\alpha}$ are approximately 15-20 eV and 3-7 eV respectively at the
observed EW$_{\rm H\alpha}$.  We can extend this calculation to
include other n=1 and n=2 transitions aside from Ly$\alpha$ and
H$\alpha$.  However, because of the greatly reduced oscillator
strengths, these lines are significantly weaker.  For instance, we
have modeled H$\beta$ EW to be $<$ 2 eV.  Rotational broadening makes
a huge impact on the observability of the Fe Ly$\alpha$ transition.
While the spectral resolution of Chandra and Astro E2 is more than
adequate to detect such a line, the Ly$\alpha$ line will be shallow
and difficult to detect. We expect rotational broadening to be a small
effect for P$\alpha$, however, the redshifted energy of the
P$\alpha$ transition of 330 eV would make it difficult due to
interstellar absorption.   

In calculating these profiles of Stark broadened lines, we have
assumed a pure hydrogen atmosphere.  However, the Stark broadening
scale given in equation (\ref{eq:StarkScale}) depends on the
background composition.  Namely $\Delta E_{\rm Stark} \propto
\left(2A/(Z+1)\right)^{2/3} Z^{1/3}$.  For solar composition, $\Delta
E_{\rm Stark}$ is larger by 14\%.  This reduces the required $N_{\rm
Fe}$ needed to reproduced the observed EW by 10\%.  For a pure He
atmosphere $\Delta E_{\rm Stark}$ is larger by a factor of 2.42, which
reduces the required $N_{\rm Fe}$ to reproduce the observed EW by
60\%.  For an atmosphere that consist of equal parts He and H by
number, we find the required $N_{\rm Fe}$ is reduced by about 50\%.
Such He-rich atmosphere would demand accretion from a H-poor donor
such as 4U 1820-30, but the initial spallation would reduce the He
fraction at the photosphere.

The formation of the He-like Fe spectral feature in the
low-temperature part of the burst is an issue that warrants further
study.  In Saha equilibrium, H-like Fe transitions to He-like Fe at
1.2 keV (CPM).  We note however that the microphysics of this line is
significantly different from the H$\alpha$ line.  Its instrinsic
broadening is due to thermal Doppler effects because the lack of
degenerate opposite parity states suppresses the linear Stark effect
(Bethe \& Salpeter 1957). Modeling this feature requires a separate
calculation which is beyond the scope of this paper.

If accretion/spallation is indeed not the endpoint of accretion onto a
neutron star, we expand on radiative levitation as a means to suspend
the Fe, which we first pointed out in BCP.  We give a simple estimate
for the Fe column that could be suspended by radiative levitation.  We
find that radiative levitation could suspend a column of Fe needed to
produce the line.  However, this result depends sensitively on the
exact nature of the background spectra.  Further work will be
necessary as the continuum spectra could be much harder than the
blackbody background that we have assumed.  

Though we assume solar metallicity throughout our calculations, the
composition of the accreting material remains highly uncertain.  If
the metallicity of the flow is sufficiently supersolar, the disparity
between the accretion/spallation Fe column and that implied from
observation can be resolved.  However, the contrast between the deep
dips and the unobscured state of \EXO\ indicates that the metallicity
may be subsolar by a factor of 2-7 (Parmar et al. 1986).  More recent
work has found that the abundance ratios of Mg, Ne and O are
consistent with solar with large uncertainties (Jimenez-Garate, Schulz
\& Marshall 2003).  The intrinsic metallicity of the flow remains
largely uncertain.

\acknowledgements

We thank the anonymous referee and Deepto Chakrabarty for a careful
reading of our manuscript.  Their comments greatly clarified this
work. We thank R. Sunyaev for detailed discussions and for
highlighting the importance of the fine structure splitting for high Z
elements.  We thank S. W. Davis for discussions on numerical solutions
to radiative transfer problems.  This work was supported by the
National Science Foundation (NSF) under PHY 99-07949, and by the Joint
Institute for Nuclear Astrophysics through NSF grant PHY 02-16783.
Support for this work was also provided by NASA through Chandra Award
Number GO4-5045C issued by the Chandra X-ray Observatory Center, which
is operated by the SAO for and on behalf of NASA under contract
NAS8-03060. I.W. receives partial support from NSF grant AST-0307273.

\end{document}